\documentclass[a4paper,fleqn]{cas-sc}
\usepackage{url} 
\usepackage{lineno}
\usepackage{soul}
\usepackage{multirow}
\usepackage{amsmath}
\usepackage{rotating}
\usepackage{gensymb}
\usepackage[authoryear]{natbib}

\def\tsc#1{\csdef{#1}{\textsc{\lowercase{#1}}\xspace}}
\tsc{WGM}
\tsc{QE}

\begin{document}
\let\WriteBookmarks\relax
\def\floatpagepagefraction{1}
\def\textpagefraction{.001}

\shorttitle{Toward a digital twin of the Great Barrier Reef}    

\shortauthors{Hill et. al.}  

\title [mode = title]{Toward a digital twin of the Great Barrier Reef: impact of extreme model resolution on tidal simulations}

\author[1]{Jon Hill}[orcid=0000-0003-1340-4373]
\cormark[1]

\author[2,3]{Ana Vila-Concejo}[orcid=0000-0003-4069-3094]
\author[1]{Katherine C. Lee}[orcid=0000-0002-8922-3926]

\affiliation[1]{organization={Department of Environment and Geography, University of York},
                country={UK}}
\affiliation[2]{organization={Geocoastal Research Group, School of Geosciences, University of Sydney},
                country={Australia}}
 \affiliation[3]{organization={Marine Studies Institute, Faculty of Science, University of Sydney},
                country={Australia}}

\begin{highlights}
\item Digital twins are a future research direction that can utilise extremely high resolution data and may incorporate numerical models.
\item Only extremely high resolution (5 m) numerical models of tidal processes can accurately capture the tidal filling and emptying of lagoons. 
\item Tidal dynamics are dependent on model resolution which has implications on simulating larval dispersal in future digital twins.
\end{highlights}

\begin{abstract}
Coral reefs are topologically complex environments with a large variation over small spatial-scales. The availability of high resolution data (metre-scale) to study these environments has increased rapidly such that many researchers are actively engaged in creating a `digital twin' of these environments to aid protection and management. However, as with any model, a digital twin will only be as useful as the data used to create it. Previous numerical modelling work on coral reefs has been carried out at a range of resolutions from 10s to 1000s of metres, but to date there has been no comprehensive study on the impact of extreme model resolution at metre-scale. Here, we simulate the Capricorn Bunker region of the GBR in a high resolution, multi-scale model using grid scales of 20,000 m to 5 m and compare that to the models with minimum grid scales of 250 m and 50 m. It is shown that the observable physical processes are best simulated at extremely high resolutions, though the intermediate resolution model performs well also. The low resolution model, whilst using a resolution comparable to a number of previous studies, does not sufficiently capture local-scale processes. Numerical models play a vital role in creating a digital twin of coastal seas as they contain the mathematical representation of the biophysical and chemical processes present but are currently at a coarser resolution than satellite and bathymetric data on which digital twins could be based. Bridging this resolution gap remains a challenge.

\end{abstract}

\maketitle

\section{Introduction} 

The recent increase of high-resolution bathymetric data, including derived from LiDAR surveys, opens up a number of opportunities for refining coastal ocean models \citep{Kutser2020-xr}. This advancement of high resolution data holds the potential to begin constructing `digital twins' of coastal environments \citep{Tzachor2023-nk}. Such digital twins, capable of mirroring the real-world dynamics of coastal process with two-way automated data flow between the numerical (digital) and the real versions, would be valuable tools for informed planning and management of human activities within the coastal zones \citep{Hazeleger2024-jl}. The flow of information between these digital representations and their real-world counterparts could be facilitated through automated extraction of flow patterns from high-resolution satellite imagery, such as PlanetScope's 3 x 3 m resolution \cite[e.g.][]{Tlhomole2025-tv}, enabling both instantaneous model validation and continuous data assimilation for performance tuning, for example. The application of such digital twins includes assessing the environmental impacts of anthropogenic activities and management of Marine Protected Areas, especially in sensitive areas, such as coral reefs \citep{Durden2025-ec}.

Coral reef ecosystems present a challenge for numerical modelling and creating a digital twin. These environments are inherently complex, exhibiting a large degree of heterogeneity across various spatial scales \citep{Torres-Pulliza2020-sb}. Accurately simulating hydrodynamic processes, such as tidal propagation, is challenging within such heterogenous settings and has long known to be so \citep{Bode1997-ds}. Therefore, the resolution employed, both in the model and of any associated data, becomes important in order to capture this spatial heterogeneity \citep{Saint-Amand2023-dg}. Previous studies have shown tidal phenomenon have clear impacts of both coral and fish larval dispersal and retention. The flow patterns within these ecosystems directly influence a number of ecological processes, including coral larval dispersal \citep{Andutta2012-he}, fish larval dispersal \citep{Burgess2007-go, Booth2000-fd}, nutrient transport \citep{Wolanski1988-nu}, and the distribution of dissolved oxygen \citep{Gruber2017-ki}. A number of studies have examined the role tidal currents have played on coral larval retention on the Great Barrier Reef \citep{Andutta2012-he, Wolanski2024-qi}. In particular, the phenomenon of `sticky waters' where mesoscale eddies trap larvae behind reefs \citep{Wolanski2000-oa} are thought to be a key factor. Other studies show that this is a function of a dense reef matrix \citep{Andutta2012-he}. The formation of tidal eddies have been implicated as a major driver of fish larval retention and dispersal on the GBR \citep{Booth2000-fd, Burgess2007-go}. In particular, `phase eddies' \citep{Black1987-cb} where the unsteady tidal flow generates eddies in the lee of islands at both the flood and ebb phases of the tidal cycle, are thought to be major factor of retention of passive larva and eggs by trapping them in the eddy such that the reversal of flow bring them back over the parent reef \citep{Burgess2007-go}. Accurate models of hydrodynamics are therefore essential for predicting the connectivity between reef habitats, informing conservation strategies aimed at maintaining genetic diversity and promoting reef resilience \citep{Elmhirst2009-wu, Williamson2016-px}.

Robust hydrodynamic models, and hence digital twins, can therefore serve as a vital tool for bridging the gap between physical processes and ecological outcomes, providing crucial insights on these ecologically significant ecosystems \citep{Tzachor2023-nk}. Previous investigations have demonstrated the necessity of resolutions ranging from approximately 250 m to 500 m to adequately represent ocean and tidal currents in reef areas \citep{Saint-Amand2023-dg}, which can then explain differences in high resolution tidal simulations \citep[e.g.][]{Mawson2022-ni} vs. lower resolution simulations \citep[e.g.][]{Harker2019-oz}. Model resolution has long been known to be a fundamental issue in representing physical processes in ocean and coastal models, with unstructured meshes one of the possible solutions \citep{Greenberg2007-cd}. However, studies have yet to quantify the impact of ultra-high resolution simulations, specifically those operating at sub-100 m scales, which is the resolution that satellite imagery and bathymetric data can new be readily obtained \citep{McCarthy2022-ov}, though models are being developed which can simulate at this scale \citep{Wagner2025-no}. This resolution gap is critical, as the scale at which processes are resolved directly influences the accuracy of predictions, particularly when assessing environmental change or understanding secondary processes, be they large-scale phenomena like sea-level rise, localized interventions such as coastal engineering projects, or assessing ecological impacts, such as larval dispersal. 

Before creating digital twins of complex heterogenous environment such as coral reefs, there is therefore a need to evaluate the impact of extreme model resolution on hydrodynamics. In this context, `extreme' is defined as metre-scale resolution within a broader regional-scale domain. This necessitates the adoption of unstructured mesh models, which offer the flexibility to spatially vary resolution, concentrating computational effort in areas of particular interest \citep[e.g.][]{Zhang2023-kk}. Coral reefs, with their inherent spatial variability, pose a significant challenge for digital twin development, yet they are also crucial hotspots of marine biodiversity facing unprecedented threats \citep{Hoegh-Guldberg2007-ed}. The rate of coral reef degradation, with approximately 10\% already lost and a further 60\% at risk, underscores the need of developing robust predictive tools \citep{Eddy2021-gp}. These ecosystems therefore present an ideal, albeit challenging, test-bed for evaluating the impact of model resolution within a digital twin framework.

This study uses a highly multi-scale numerical model to simulate tidal dynamics within the Great Barrier Reef (GBR). We undertake a validation of the model against observational tidal gauge data across three distinct model resolutions. Subsequently, we perform a detailed comparative analysis of tidal dynamics, employing a range of quantitative metrics, focusing on a single atoll reef for which exceptionally high-resolution (0.25 m) bathymetric data is available \citep{Talavera2021-xo, Harris2023-zk}. 

\subsection{Study site}

\begin{figure}[ht]
\noindent\includegraphics[width=\textwidth]{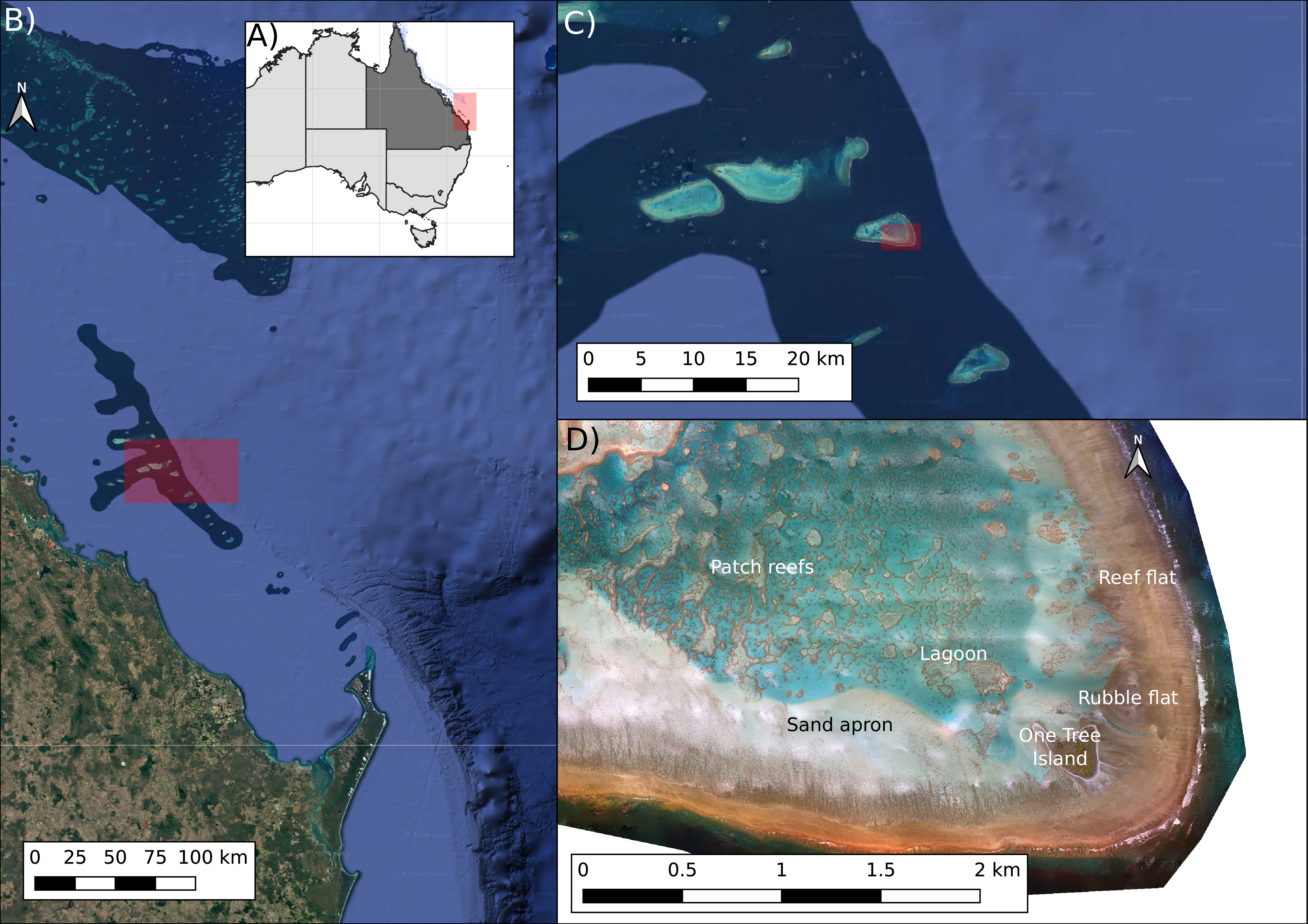}
\caption{Location of One Tree Reef, within the Great Barrier Reef's Capricorn Group islands. A) shows the Australian coastline, with blue showing the reef locations. The red square then shows the location of B). Panel B) shows the Capricorn Group of atoll islands. Panel C) shows One Tree Reef (location show by the red square in B). Panel D shows a zoom-in of One Tree Island with various geomorphological features highlighted. Panel D is the RGB part of the LiDAR data described in \citet{Talavera2021-xo, Harris2023-zk}.}
\label{fig:overview}
\end{figure}

One Tree Island (OTI) and Reef (OTR) are situated in the Capricorn Group (Fig. \ref{fig:overview}), in the southern Great Barrier Reef (GBR) \citep{Davies1976-rk}. OTR spans and area of around 40,000 square metres, including a 5.5 km by 3 km reef and a small, vegetated, shingle cay located in the southeastern corner of the platform \citep{Bauder2023-qv}. The lagoon is entirely rimmed by reef crest. The reef contains a main lagoon in the centre of the reef, comprising of depths that vary between 0.75-20 m, with smaller lagoonal areas adjacent. The main lagoon is relatively shallow, with a maximum ponded depth of around 6 m at the northern edge where  fine mud dominates the sediment. It contains a number of small patch reefs that reach the ponded surface. At the shallower southern and eastern edges, the lagoonal sediment is dominated by coarse sand. \citep{Kosnik2015-gc}. The tidal dynamics at One Tree are governed by the semidiurnal principal lunar (M$_2$) and solar (S$_2$) components with a diurnal divergence (ponding effect) as the lagoon is separated from the ocean \citep{Wilson1985-xm} by the reef crest. The island therefore experiences a primary tidal pattern that comprises two high and low tides per day, with a tidal range of approximately 3.4 m \citep{Hatcher1985-tf}, but becomes almost diurnal in nature during neap periods. The tidal patterns causes the reef platform encompassing OTR to be exposed for 5-6 hours at low spring tide which means the lagoon has limited exchange with the open ocean for significant parts of the tidal cycle \citep{Ludington1979-mn, Wilson1985-xm, Harris2014-sj}. 

Previous research investigating physical processes at One Tree have focused on wave energy dissipation \citep{Duce2022-cd, Perris2024-ot}; sediment transport and geomorphology \citep{Harris2014-sj, Talavera2021-xo, Vila-Concejo2022-wk}; wind-driven circulation \citep{Frith1986-dx, Burgess2007-go}; and tides \citep{Wilson1985-xm, Ludington1979-mn}, but no study has simulated tides with high resolution bathymetric data.

\section{Methods}
\subsection{Model details and set-up}

Coastal models that represent coastal and oceanic regions can be used to predict any alterations in tidal cycles due to changes in infrastructure, coastal alterations or sea level change \citep[e.g.][]{Baker2020-tg, Lee2022-ny, Mawson2022-ni}. In this work, \emph{thetis} was employed, which is a finite element based coastal ocean model that implements both 2D and 3D equations \citep{Karna2018-lv}. \emph{thetis} solves the non-conservative form of the non-linear shallow water equations: 

\begin{equation}
A(H)\frac{\partial \eta}{\partial t} + \nabla \cdot (\tilde{H} \mathbf{u}) = 0,
\label{eq:swe1}
\end{equation}

\begin{equation}
\frac{\partial \mathbf{u}}{\partial t} + \mathbf{u} \cdot \nabla \mathbf{u} + f \mathbf{u}^{\perp}  + g \nabla \eta + \nabla\frac{p_a}{\rho_0} = -\frac{\tau_b}{\rho \tilde{H}} + \nabla \cdot ( \nu (\nabla \mathbf{u} + \nabla \mathbf{u}^T)),
\label{eq:swe2}
\end{equation}

\noindent where $A(H)$ is the wetting and drying formulation (see below), $\eta$ is the water elevation, $H_d$ is the total water depth, $\nu$ is the kinematic viscosity of the fluid, and $\mathbf{u}$ is the depth-averaged velocity vector. The Coriolis term is represented as $f\mathbf{u}^{\perp}$, where  $\mathbf{u}^{\perp}$ the velocity vector rotated counter-clockwise over $90^o$. In turn, $f = 2 \Omega \textnormal{sin}\zeta$ with $\Omega$ corresponding to the angular frequency of the Earth's rotation and $\zeta$ the latitude. Note that the continuity equation (eq. \ref{eq:swe1}) has modified following \citet{Karna2011-af} such that $\frac{\partial \eta}{\partial t} + \frac{\partial \tilde{h}}{\partial t} = A(H)\frac{\partial \eta}{\partial t}$, where $\tilde{h}$ is defined by:

\begin{equation}
\begin{aligned}
    \tilde{h}(\eta,h) & = h + f(H), \\
    f(H) & = \frac{1}{2}(\sqrt{H^2 + \alpha^2} - H)
\end{aligned}
\label{fluct_bathy}
\end{equation}

\noindent such that during low tide, areas of the reef will be exposed. Here, $f(H)$ ensures redefined total depth remains positive, via the wetting and drying parameter ($\alpha$), $\eta$ is the free surface height and $h$ is the original static bathymetry depth. Bed shear stress ($\tau_b$) effects are represented through the Manning's $n$ formulation as:

\begin{equation}
\frac{\tau_b}{\rho} = g n^2 \frac{|\mathbf{u}|\mathbf{u}}{{H_d}^\frac{1}{3}}
\label{eq:bss}
\end{equation}

For time-stepping, a second-order DIRK22 discretisation was used with a constant time step which varied depending on the smallest grid-scale used (see below).

The wetting and drying $\alpha$ parameter was set as a function of element size and slope and could vary between 0.1 and 75. The smallest values are used where the mesh resolution is high and the slope is low and the highest values used where the mesh is coarse and the slope is high. The models were implemented following the Galerkin finite element discretisation (DG-FEM), using the P1DG-P1DG velocity-elevation finite element pair \citep{Angeloudis2018-jl, Baker2020-tg}. Discretised equations are solved using a Newton nonlinear solver algorithm via the PETSc library \citep{balay2001petsc}.

\subsection{Mesh Generation and Model Construction}

The mesh needs to be of sufficient extent to avoid boundary condition issues, whilst containing enough resolution to resolve any areas of complex bathymetry or coastal geomorphology, but also, in this case, the high resolution bathymetric data on OTR. As \emph{thetis} is a finite-element model it can easily use spatially-varying mesh resolution, allowing higher resolution near areas of interest, with coarse resolution away from the site \citep[e.g][]{Hill2023-pm}. Here, we created three meshes which are identical except the edge-length metric employed in the area around OTR. Bathymetric and topographic data came from two sources. The GBR bathymetric data of \citet{Beaman2010-no}, and the LiDAR survey carried out over OTR as described in \citet{Talavera2021-xo} and \citet{Harris2023-zk}. The \citet{Beaman2010-no} data were resampled to 400 m and 100 m resolution in UTM56 projection space. The OTR data was resampled to 5 m resolution in UTM56 projection space, with small amounts of missing data filled in via linear interpolation. 

The mesh boundaries are created from bathymetric/topographic data by creating a contour at a set height to create the `landward' boundary, and joining that via a smooth arc to create the open boundary of the model. The contour used here is the +5 m contour. The contour was then modified by removing small islands of less than 3000 m circumference. The contour was also modified to remove narrow bays or other coastal features that contains fewer than two elements width. An external forced boundary was then added between -22.53\degree latitude, 150.78\degree longitude at the northern extent to -25.93\degree latitude, 153.18\degree longitude on the southern edge, going out around 150 km from the modern coastline. Processing and contour creation were carried in QGIS \citep{QGISorg2020-ro}. GIS data were converted to meshes via \emph{qmesh} \citep{Avdis2018-wy} and \emph{gmsh} \citep{Geuzaine2009-dd}.

Mesh resolution for the model was determined by five factors: water depth from the \citet{Beaman2010-no} data, $\mathcal{H}_b$; water depth from the OTR data, $\mathcal{H}_o$; distance from the `landward' boundary, $\mathcal{H}_{lb}$; distance from forced boundary $\mathcal{H}_{ob}$; and the distance from OTR, $\mathcal{H}_{dm}$ or $\mathcal{H}_{dh}$. The final mesh resolution, $\mathcal{H}$, is the minimum of the factors:

\begin{equation}
    \mathcal{H} = \mathrm{min}\left(\mathcal{H}_b,\mathcal{H}_o,\left[ \mathcal{H}_{dh} \middle| \mathcal{H}_{dm}\right],\mathcal{H}_{lb},\mathcal{H}_{ob} \right)
\end{equation}

Bathymetric/topographic control was governed by:

\begin{equation}
    \mathcal{H}_b = 
        20000 \frac{\exp \left( \frac{\left( -b - 165 \right )}{50} \right)}{\exp \left( \frac{\left( -b - 165 \right )}{  50} \right) +1}
\end{equation}

\noindent where $b$ is the topographic height. A different form was used with the high resolution bathymetric data of OTR:

\begin{equation}
    \mathcal{H}_o = 0.4b^2+0.1b+5
\end{equation}

Resolution as a function of distance from OTR was governed by creating a polygon over OTR where distance was set to zero. The \emph{gdal} proximity function \citep{GDAL/OGR-contributors2025-mn} was then used to create a raster which contained the distance (in metres) from that polygon area. Mesh resolution was then calculated using:

\begin{equation}
    \mathcal{H}_{dh}= \frac{D}{6} + 20.0
\end{equation}

\noindent where $D$ is the distance from the polygon. For the medium resolution mesh, $\mathcal{H}_d$ was:

\begin{equation}
    \mathcal{H}_{dm}= \frac{D}{6} + 50.0
\end{equation}

\noindent and for the low resolution mesh, this metric was not included.

The final two resolution metrics were constructed in the same way, but using different parameters. For the `landward' boundary resolution, $\mathcal{H}_{lb}$, increased from 500 m to 20 km linearly to a distance of 50 km with the first 1000 m away from the boundary fixed at 500 m resolution. For the open boundary the resolution, $\mathcal{H}_{ob}$, increased from 2 km at the boundary to 20 km away from the boundary at a distance of 50 km, with the nearest 6 km fixed at 2 km resolution.

\begin{figure}[ht]
\noindent\includegraphics[width=\textwidth]{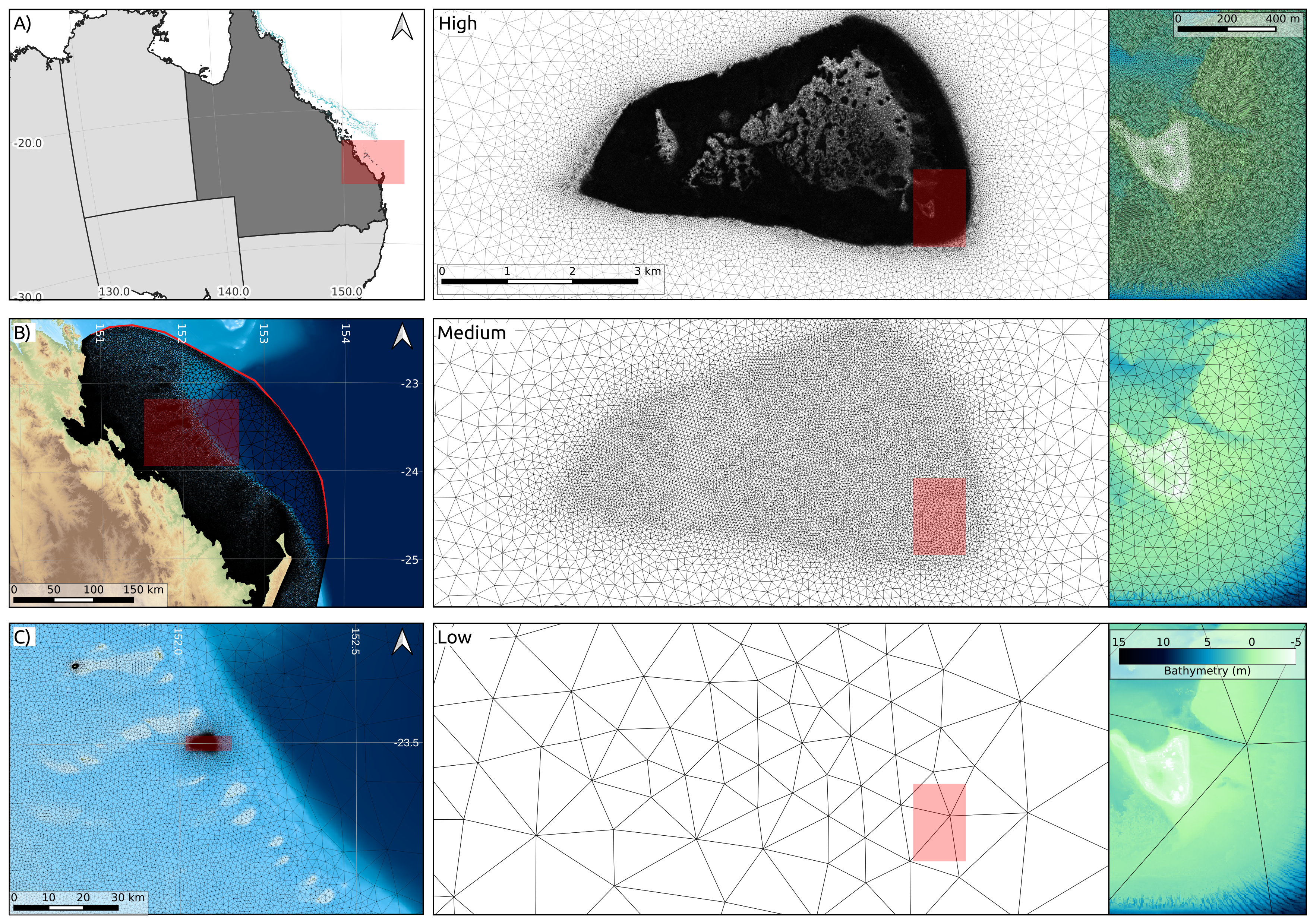}
\caption{Overview of the model domain and mesh (A and B), with close up of the OTR areas (C). The modelled panels show the mesh for the HR (top), MR (centre) and LR (bottom) simulations around OTR, with a close-up around the One Tree Island (right panels).}
\label{fig:mesh}
\end{figure}

Mesh resolution therefore varies from 20 km in the Pacific Ocean to either 5 m, 50 m or around 250 m around OTR (Fig. \ref{fig:mesh}). The three meshes can were then constructed using the correct components of the overall mesh metric (Table \ref{tab:mesh_metric}). The meshes contained between 1.1 million 100,000 elements. Henceforth, the three meshes will be referred to as HR (high resolution), MR (medium resolution) and LR (low resolution).

\begin{table}[ht]
\label{tab:mesh_metric}
\caption{Definition of the final mesh metric used in the three meshes for low, medium and high resolution models.}
\begin{tabular}{l|l}
Mesh  & Mesh metric \\ \hline
LR    & $\mathcal{H} = \mathrm{min}\left(\mathcal{H}_b,\mathcal{H}_{lb},\mathcal{H}_{ob} \right)$ \\
MR    & $\mathcal{H} = \mathrm{min}\left(\mathcal{H}_b,\mathcal{H}_{dm},\mathcal{H}_{lb},\mathcal{H}_{ob} \right)$            \\
HR    &  $\mathcal{H} = \mathrm{min}\left(\mathcal{H}_b,\mathcal{H}_o,\mathcal{H}_{dh},\mathcal{H}_{lb},\mathcal{H}_{ob} \right)$          
\end{tabular}
\end{table}

All bathymetric/topographic datasets were interpolated onto the mesh using bilinear interpolations, with blending of different data carried out using the HRDS package \citep{Hill2019-ju}. All models used three bathymetric datasets: a 400 m version of \citet{Beaman2010-no} which covered the whole domain; a 100 m version of \citet{Beaman2010-no} covering the central part of the domain, and a 5 m version of the \citet{Talavera2021-xo} LiDAR data over OTR. All bathymetric data were projected to UTM56 coordinate reference system and use the same datum.

All models used the same numerical parameters, forcing datasets and bathymetric/topographic data. Models were forced using TXPO version 9 \citep{Egbert2002-cn} starting from November 22, 2022 at midnight UTC. The only minimal change between models (except the mesh) was a change in the model timestep from 180 s for the LR model to 90 s for the MR and HR models. Output was stored every 15 minutes for subsequent analysis. Due to the long run times required for the HR model these models were run for 16 days plus an additional 2 days spin-up. The MR and LR models were run for 30 days plus an additional 2 days spin-up. All model parameters are detailed in the supplementary information.

\subsection{Particle tracking}

In order to establish the impacts of model resolution on physical processes around OTR, offline Lagrangian particle tracking was employed as a method of estimating flushing times of the lagoon. Here, a modified version of \emph{parcels} \citep{Delandmeter2019-li} was used where, rather than using gridded input normally required, the velocity at each particle location was evaluated on the finite element function on the unstructured mesh. Linear interpolation in time was used between model outputs. Moreover, the particles were held in position when water elevation was lower than the bathymetric height so particles did not move when `dry'. A set of 10,000 random locations was generated within the OTR area and used as starting locations for all particle simulations. The particle tracking used a 4th order Runge-Kutta timestepping algorithm, with a timestep of 120 seconds. Outputs (particle positions and time) were stored every 240 seconds.

Particles were released and tracked for 48 hours. A total of 32 release times were used for each model; 16 during the spring tide period and 16 during the neap part of the tidal cycle. Release times started at 314,880 seconds (3.64 days; spring tide) or 1,076,400 seconds (12.46 days; neap tide) after the model start time and were 10,000 seconds apart until 16 releases had been carried out. To establish the flushing rate, the number of particles were counted within the OTR area at each particle tracking output step. The mean number of particles per step was then calculated across the 16 spring tide release points and the 16 neap tide release points. To compare the effects of model resolution on the lagoon flushing, two statistical tests were carried out. First, an ANOVA with Tukey post-hoc HSD was carried out on the 16 groups at 5 hour intervals. This therefore computes any statistical difference in the flushing rates at a static point in time across the three models resolutions. Second, a Functional Data Analysis ANOVA was carried out which examines statistical difference in the groups of lines as a set of whole time series. Here, we used the R package \emph{fdANOVA} \citep{Gorecki2019-ij} using a b-spline basis function. We report the results of the L2N test and the TRP (Test Random Projection) using 30 Gaussian projections with 1000 permutations each. This test produces three p-values based on different statistical tests.

The pathways for each particle were also tracked and the location of where they crossed the OTR outline was noted as an inflow or outflow point. Multiple  events were counted for each track. To count the number of tracks crossing each location a Kernel Density Estimation function using a quartic kernel shape was applied to the exit or entry points with a radius of 25 m and stored in a raster with 5 m pixel size.  

\subsection{Model Validation}

\begin{figure}[ht]
\noindent\includegraphics[width=\textwidth]{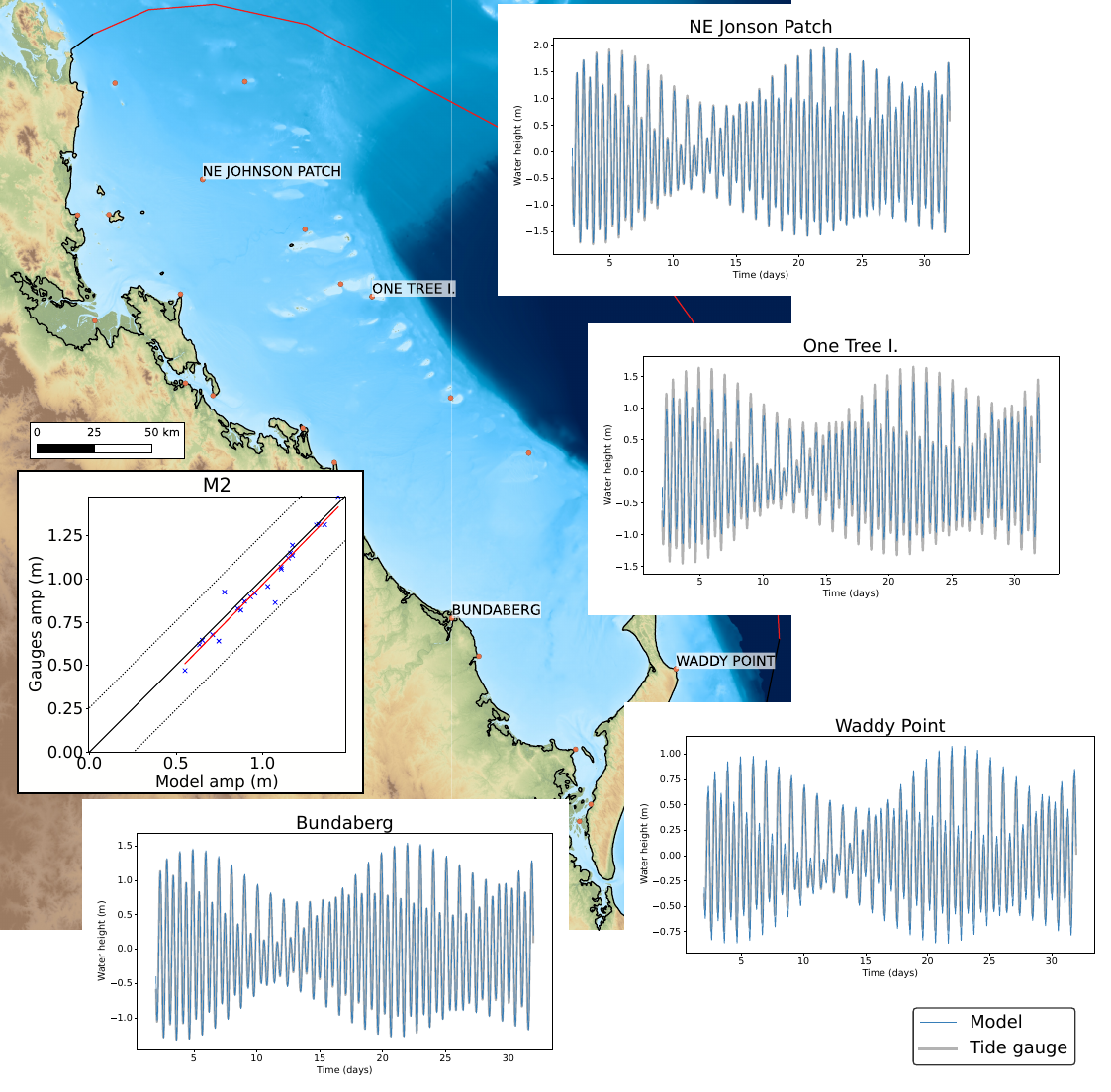}
\caption{Verification of the tidal model at medium resolution with tide gauges within the domain. The cross plot shows the
correlation of the M$_2$ amplitude of the model gauges (x-axis) against gauge data (y-axis). The red line indicates the line of best fit, with the solid black indicating the 1:1 relationship and the dotted line the standard deviation of the measured amplitudes. Four example tidal gauges are show from across the domain (see central map for locations), which show the theoretical tide at the location in grey, with the modelled tide overlain in blue.}
\label{fig:tide_gauges}
\end{figure}

To demonstrate the model is capable of replicating the tidal system in the area we validated the model against 23 tide gauges in the region using the MR and LR models; note that only the One Tree Islet tidal gauge is located in the region where the resolution is altered. The validation shows good agreement for the amplitude four major tidal components in the region (Table \ref{tab:tides}) using a mix of regression and the error metric proposed by \citet{Cummins2017-ja}. The model can accurately replicate the tidal elevation of all the tide gauges in the region (Fig. \ref{fig:tide_gauges}). The change in resolution does not alter the comparison, except for the One Tree Islet gauge station, as expected. 

\begin{table}[ht]
\caption{Various tidal model performance metrics split by tidal constituents. The first two rows are based on the error metric of \citet{Cummins2017-ja}, with relative error calculated using the mean measured amplitude. The Pearson's $r$ and p-value were calculated using the standard methods.}
\label{tab:tides}
\begin{tabular}{l|llll}
               & M$_2$       & S$_2$       & K$_1$       & O$_1$  \\ \hline
Error          & 0.0715 m    & 0.0205 m    & 0.0077 m    & 0.0061 m\\
Relative error & 7.38\%      & 5.96\%      & 2.98\%      & 4.56\% \\
Pearson's $r$  & 0.9719      & 0.9271      & 0.8212      & 0.8966 \\
p-value        & $\ll 0.001$ & $\ll 0.001$ & $\ll 0.001$ & $\ll 0.001$ \\ \hline
\end{tabular}
\end{table}

\section{Results}

\begin{figure}[ht]
\noindent\includegraphics[width=\textwidth]{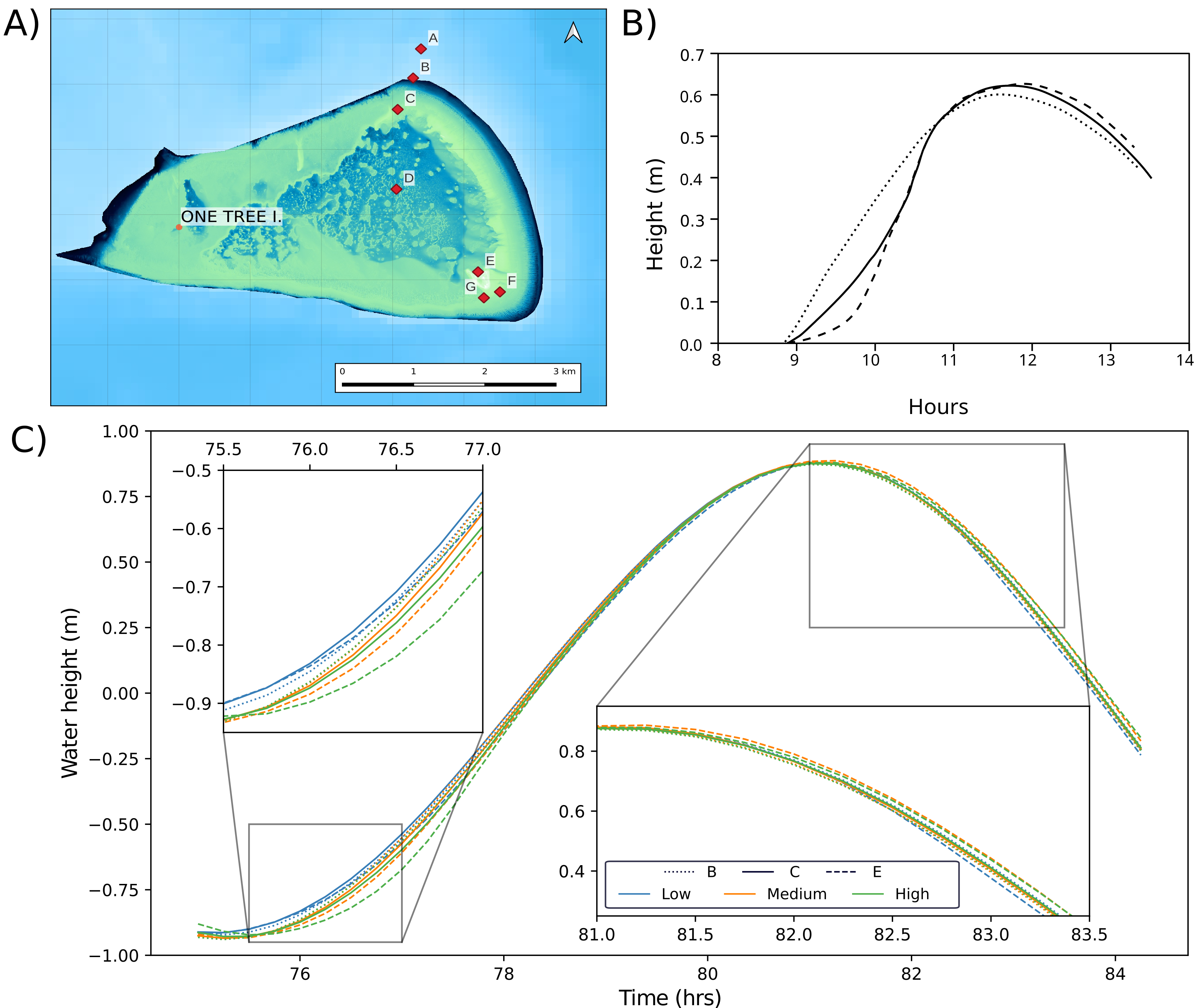}
\caption{Location of points (B, C and E) around OTR (panel A) where the ponding effect can be measured. Points are those used by \citet{Wilson1985-xm} to demonstrate the lagged filling and emptying of the lagoon at OTR (panel B; after \citet{Wilson1985-xm}). Linetype indicates the gauge location and colour denotes the model resolution. Two insets in panel C show the lag phenomenon at rising and falling tide respectively.}
\label{fig:wilson}
\end{figure}

Model resolution has a clear impact on correctly simulating the ponding effect on OTR. \citet{Wilson1985-xm} showed that there is a time lag
in water height from outside the lagoon, the northern edge of the lagoon and to the cay on the southern side and the centre of the lagoon. This results in the water elevation at the cay lagging behind the northern edge of the lagoon by up to one hour, which in turn lags behind the elevation outside the lagoon (Fig. \ref{fig:wilson}). The LR model did not capture this observed phenomenon, with the tidal elevation at the cay slightly lower than, but not out of phase of, the outer lagoon or the outer reef locations. Both the MR and HR simulations showed this phase lag, but it was much more pronounced in the HR model (Fig. \ref{fig:wilson}A). This shows the clear difference, at a local scale, of model resolution with higher resolution simulations successfully capturing the filling and emptying of the lagoon over a tidal cycle. 

\begin{figure}[ht]
\noindent\includegraphics[width=\textwidth]{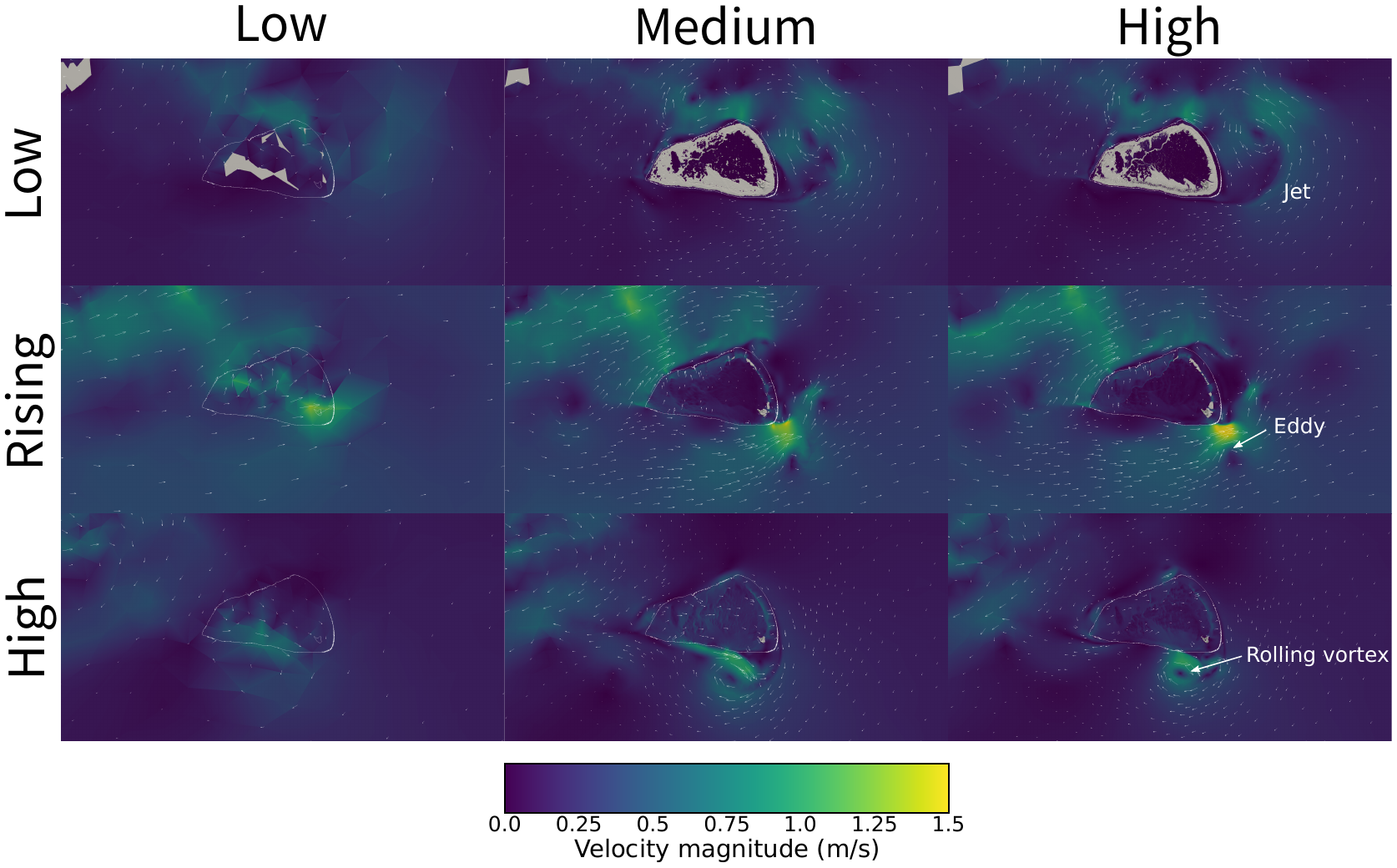}
\caption{Comparison of the spatial velocity patterns around OTR for the three model resolution (columns) at three points
in the tidal cycle (rows: low, rising and high tide) during the peak spring tide period. Arrows show directionality and are 
scaled to the magnitude, also shown by the colour scale. [JH: label features mentioned in text]} 
\label{fig:spring_vel}
\end{figure}

The differences in flow as a function of resolution are clearly seen in and around OTR. During the spring tide phase in the HR model (Fig. \ref{fig:spring_vel}) there is a clear eastward jet of fast flowing water that arcs along the southern side of OTR and then turns north on the eastern side of the reef at low tide (ebb, with predominantly easterly flow). This jet is present in all three simulations, but the flow magnitude and direction are substantially different. The LR model shows a weakly defined jet travelling to around 2 km east of the island before becoming more diffuse in a north-easterly direction. The jet increases in strength and definition as resolution increases, turning northward around 4-5 km from the island. Similarly, at rising tide a clear eddy is seen on the south-eastern corner of OTR, near OTI, which concentrates flow and produces a high velocity flow eastwards to the south of the reef rim. On the LR model this high velocity flow is concentrated on OTI itself, rather than off the southern edge of the reef, and is lower in magnitude. Finally, at high tides, the southern edge shows a series of vortices rolling along the southern edge in the HR model. These are not present in the LR model and are more diffuse in the MR model. The video in the supplementary information demonstrates a full tidal cycle for the HR model.

Further away from OTR also shows changes in flow. The top north-west corner of the plots (Heron Island) shows subtle differences despite the resolution of the models being very similar this distance from OTR. Similarly, the southern edges of all plots show subtle differences in flow across resolution as a direct consequence of differences in the flow on the southern edge of OTR, whilst the resolution here is also similar in all models.

\begin{figure}[ht]
\noindent\includegraphics[width=\textwidth]{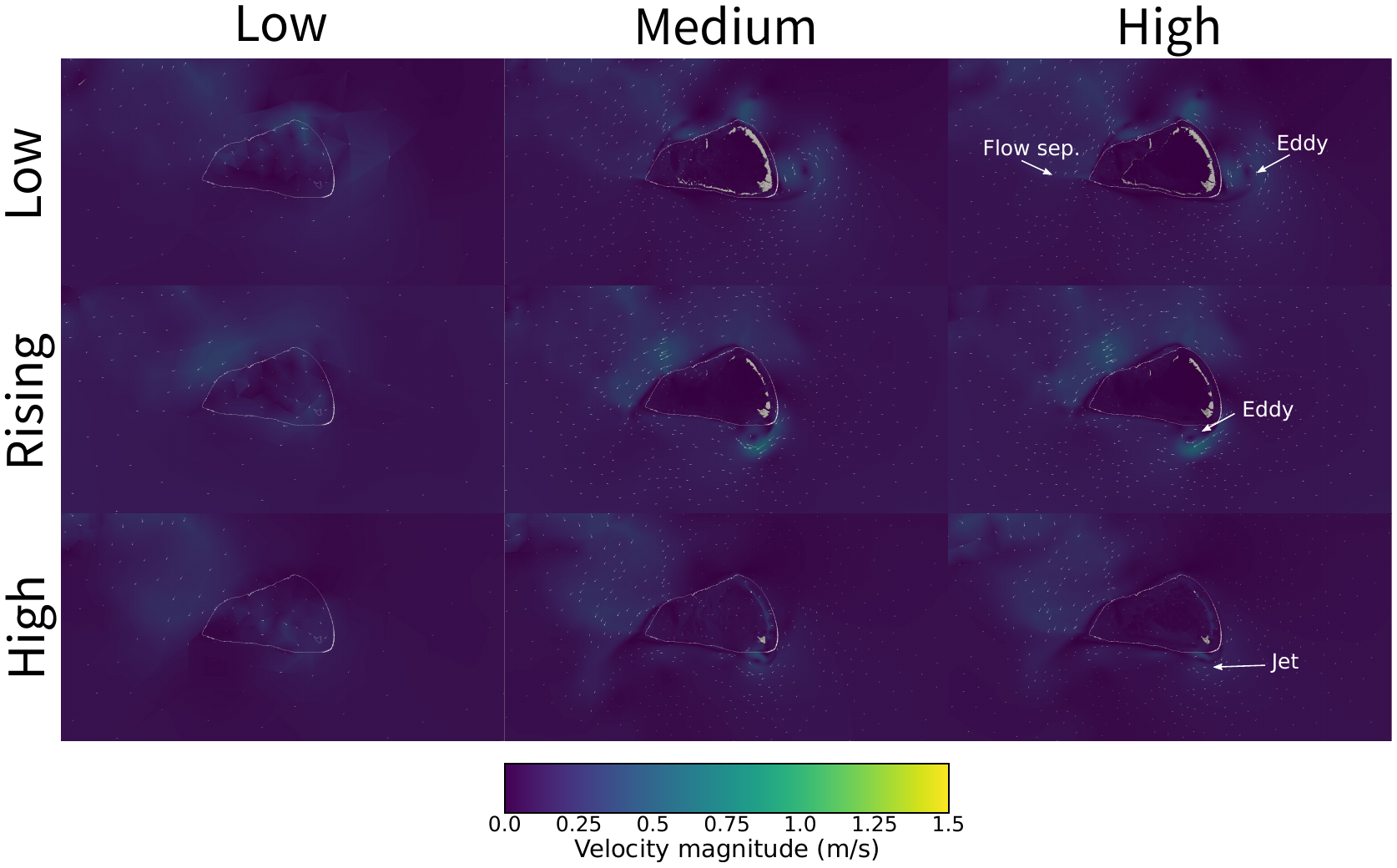}
\caption{As Fig. \ref{fig:spring_vel} but for the neap phase of the tidal cycle.  [JH: label features mentioned in text]}
\label{fig:neap_vel}
\end{figure}

During the neap tidal period there are similar, but more subtle differences in the flow field around OTR. As with the spring tidal period, there is a clear jet and subsequent vortex shedding off the southern edge of the island in an easterly direction during low tide (Fig. \ref{fig:neap_vel}). This appears completely absent in the LR simulation. Similarly, the western edge of OTR up toward Heron island is very similar in the MR and HR simulations with a clear flow separation visible, but is absent in the LR simulation. During rising tide (flood, with predominantly westerly flow) there are high magnitudes of flow in the lagoon for the LR simulation and a lack of eddies on along the southern edge of the reef, compared to the MR and HR simulations. At high tide there are substantial differences in the flow magnitude within the lagoon, with the LR simulation showing flow of around 0.5 m/s higher than the other two simulations. The LR simulation also lacks a clear jet structure from the western edge of the island, which is visible in the other simulations. 

\begin{figure}[ht]
\noindent\includegraphics[width=\textwidth]{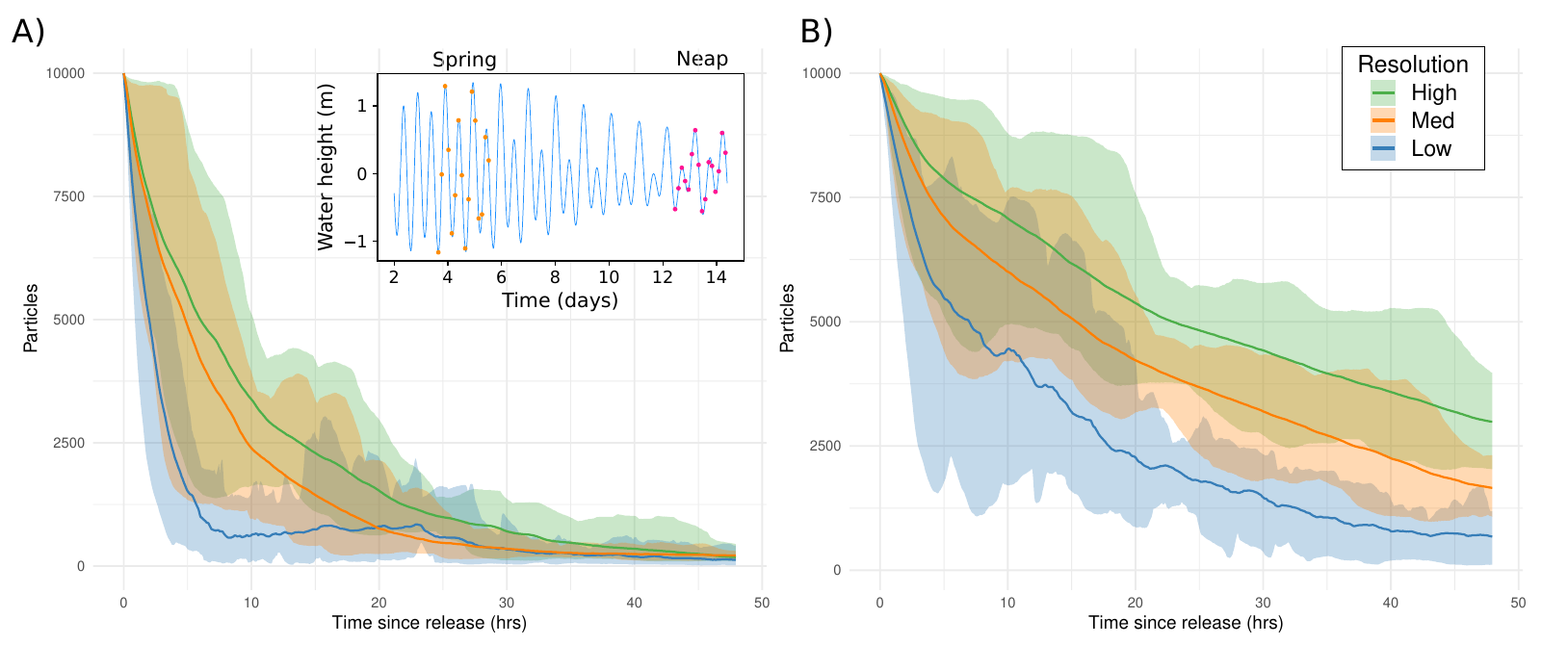}
\caption{Number of particles within the OTR area from time of release (10,000 particles) over a 48 hour period for each release (shaded areas) and the mean (solid line) for the spring tidal period (A) and neap (B). Colour denotes the model resolution. The inset graph shows the release times against model elevation - orange dots denote spring tide releases and red show neap tide releases. }
\label{fig:lagoon_flusing}
\end{figure}

The differences in flow regime then impact how the water within the lagoon is exchanged with the open ocean. Counting the number of particles that are within the lagoon area shows that the LR simulations take around 5 days to go from 10,000 particles to 1,000 during the spring tide phase (Fig. \ref{fig:lagoon_flusing}A). In contrast the MR simulations take around 15 days on average and the HR simulations 23 days. The variation in counts is also higher in the MR and HR simulations and some of these simulations show substantial increases in particles within the lagoon at times. Animations of the particle experiments show that the LR simulation particles stay in relatively coherent groups after the initial flushing from the lagoon (see supplementary information). There are only a small number of particles left in the lagoon after the first few hours. In contrast the HR simulation shows that particles remain within the low velocity area of the lagoon, flushing out along specific pathways. Flow is severely inhibited within the lagoon, meaning particles stay there for a much longer period in time with flushing into the open ocean further inhibited by the low tide creating a barrier. The MR simulation shows a lot of similarities with the HR simulation, but higher flow in the initial stages leads to an immediate flushing of a large number of particles. However, this is clearly a function of the release point as over the 16 spring tide releases the mean initial loss of particles is very similar to the high resolution simulation (Fig. \ref{fig:lagoon_flusing}A).

The neap period shows a much decreased flushing rate in all simulations, but similar differences are seen between the three model resolutions. The LR run takes until around 31 hours to go from 10,000 to 1,000 particles. Neither the MR or HR simulations reach 1,000 particles in the lagoon over the 2 day particle tracking experiment. The LR simulations show much greater variability (shaded area, Fig. \ref{fig:lagoon_flusing}) in flushing rates for the first 20 hours compared to the spring tide simulations, whereas the MR and HR simulations show similar variation as a function of release time. Videos in the supplementary information show the particle through time for each 2 day experiment. 

The flushing rates compared over the 16 simulations of neap and spring tide respectively, show different impacts of resolution on the flushing rate. The three model resolutions show significant differences ($p < 0.001$) for the majority of the neap tidal cycle, with MR-HR demonstrating no statistically significant differences for the the first 10 hours (Table 3). The next 10 hours show significant differences at $p<0.05$ (15 hours) and $p<0.01$ (20 hours) and all subsequent hours are significantly different to $p<0.001$. Similar MR-LR are significantly different to $p<0.01$ for hours 5 and 10, before being significantly different to $p<0.001$ for the remainder of the tidal cycle. These results are confirmed by the functional ANOVA analysis, with statistically significant differences between all three models using both test methods (Table 4). The spring tidal cycles is more complex. The MR-LR comparison shows statistically significant differences ($p<0.001$) for the first 10 hours, but then all comparisons are not statistically significantly different for the remainder of the tidal cycle (Table 3). However, the functional ANOVA results show statistically significant differences for all tests for the MR-LR comparison (Table 4). The HR-LR comparison is significantly different (to varying p-values) throughout the cycle, except for the final time (45 hours). The functional ANOVA shows a significant difference for all tests. Finally, the MR-HR resolution difference shows no significant difference at hour 5, followed by significant difference (at varying p-values) for hours 10 to 35, followed by no significant differences for the final two time outputs (Table 3). The functional ANOVA shows a mix of significant (TRP tests to $p<0.05$) and non-significant (L2N test) results (Table 4).

\begin{table}[ht]
\small
\label{tab:anova}
\caption{Difference in the mean values of the number of particles within OTR at 5 hourly intervals. The model comparison is in the 
leftmost column, i.e. `LR-HR' denotes the high resolution model mean was subtracted from the low resolution model mean. Colour intensity
indicates significance threshold: white is not significant, lightest green is $p<0.05$, middle green is $p<0.01$ and darkest green is
$p<0.001$.}
\begin{tabular}{l|lllllllll}
\textbf{Neap}   & 5  & 10  & 15  & 20  & 25  & 30   & 35  & 40   & 45    \\ \hline
LR-HR           & \cellcolor[HTML]{009901}{\color[HTML]{FFFFFF} -2403.9} & 
                  \cellcolor[HTML]{009901}{\color[HTML]{FFFFFF} -2625.3} & 
                  \cellcolor[HTML]{009901}{\color[HTML]{FFFFFF} -2983.6} & 
                  \cellcolor[HTML]{009901}{\color[HTML]{FFFFFF} -2625.3} & 
                  \cellcolor[HTML]{009901}{\color[HTML]{FFFFFF} -2625.3} & 
                  \cellcolor[HTML]{009901}{\color[HTML]{FFFFFF} -2625.3} & 
                  \cellcolor[HTML]{009901}{\color[HTML]{FFFFFF} -2897.6} & 
                  \cellcolor[HTML]{009901}{\color[HTML]{FFFFFF} -2625.3} & 
                  \cellcolor[HTML]{009901}{\color[HTML]{FFFFFF} -2446.0} \\
MR-HR          & -774.6                                                 & 
                  -1071.6                                                & 
                  \cellcolor[HTML]{9AFF99}-1100.7                        & 
                  \cellcolor[HTML]{34FF34}-1150.2                        & 
                  \cellcolor[HTML]{009901}{\color[HTML]{FFFFFF} -1143.7} & 
                  \cellcolor[HTML]{009901}{\color[HTML]{FFFFFF} -1225.3} & 
                  \cellcolor[HTML]{009901}{\color[HTML]{FFFFFF} -1246.3} & 
                  \cellcolor[HTML]{009901}{\color[HTML]{FFFFFF} -1333.4} & 
                  \cellcolor[HTML]{009901}{\color[HTML]{FFFFFF} -1363.3} \\
MR-LR          & \cellcolor[HTML]{34FF34}1629.3                         & 
                  \cellcolor[HTML]{34FF34}1629.3                         & 
                  \cellcolor[HTML]{009901}{\color[HTML]{FFFFFF} 1629.3}  & 
                  \cellcolor[HTML]{009901}{\color[HTML]{FFFFFF} 1954.9}  & 
                  \cellcolor[HTML]{009901}{\color[HTML]{FFFFFF} 1891.4}  & 
                  \cellcolor[HTML]{009901}{\color[HTML]{FFFFFF} 1730.4}  & 
                  \cellcolor[HTML]{009901}{\color[HTML]{FFFFFF} 1651.3}  & 
                  \cellcolor[HTML]{009901}{\color[HTML]{FFFFFF} 1467.5}  & 
                  \cellcolor[HTML]{009901}{\color[HTML]{FFFFFF} 1082.7}  \\ \hline
\multicolumn{1}{c}{\textbf{Spring}}    \\ \hline
LR-HR         & \cellcolor[HTML]{009901}{\color[HTML]{FFFFFF} -4057.9} & 
                  \cellcolor[HTML]{009901}{\color[HTML]{FFFFFF} -2754.3} & 
                  \cellcolor[HTML]{009901}{\color[HTML]{FFFFFF} -1550.9} & 
                  \cellcolor[HTML]{34FF34}-724.3                         & 
                  \cellcolor[HTML]{009901}{\color[HTML]{FFFFFF} -410.5}  & 
                  \cellcolor[HTML]{34FF34}-365.1                         & 
                  \cellcolor[HTML]{34FF34}-217.1                         & 
                  \cellcolor[HTML]{9AFF99}-159.7                         & 
                  -97.1                                                  \\
MR-HR         & -705.9                                                 &
                 \cellcolor[HTML]{34FF34}-987.5                         & 
                 \cellcolor[HTML]{9AFF99}-853.5                         & 
                 \cellcolor[HTML]{34FF34}-768.1                         & 
                 \cellcolor[HTML]{009901}{\color[HTML]{FFFFFF} -521.9}  & 
                 \cellcolor[HTML]{34FF34}-352.9                         & 
                 \cellcolor[HTML]{34FF34}-204.7                         & 
                 -108.1                                                 & 
                 -17.8                                                  \\
MR-LR         & \cellcolor[HTML]{009901}{\color[HTML]{FFFFFF} 3351.9}  & 
                  \cellcolor[HTML]{009901}{\color[HTML]{FFFFFF} 1766.8}  & 
                  697.4                                                  & 
                  -43.9                                                  & 
                  -111.4                                                 & 
                  12.1                                                   & 
                  12.4                                                   & 
                  51.6                                                   & 
                  79.4                                                  
\end{tabular}
\end{table}

\begin{figure}[ht]
\noindent\includegraphics[width=\textwidth]{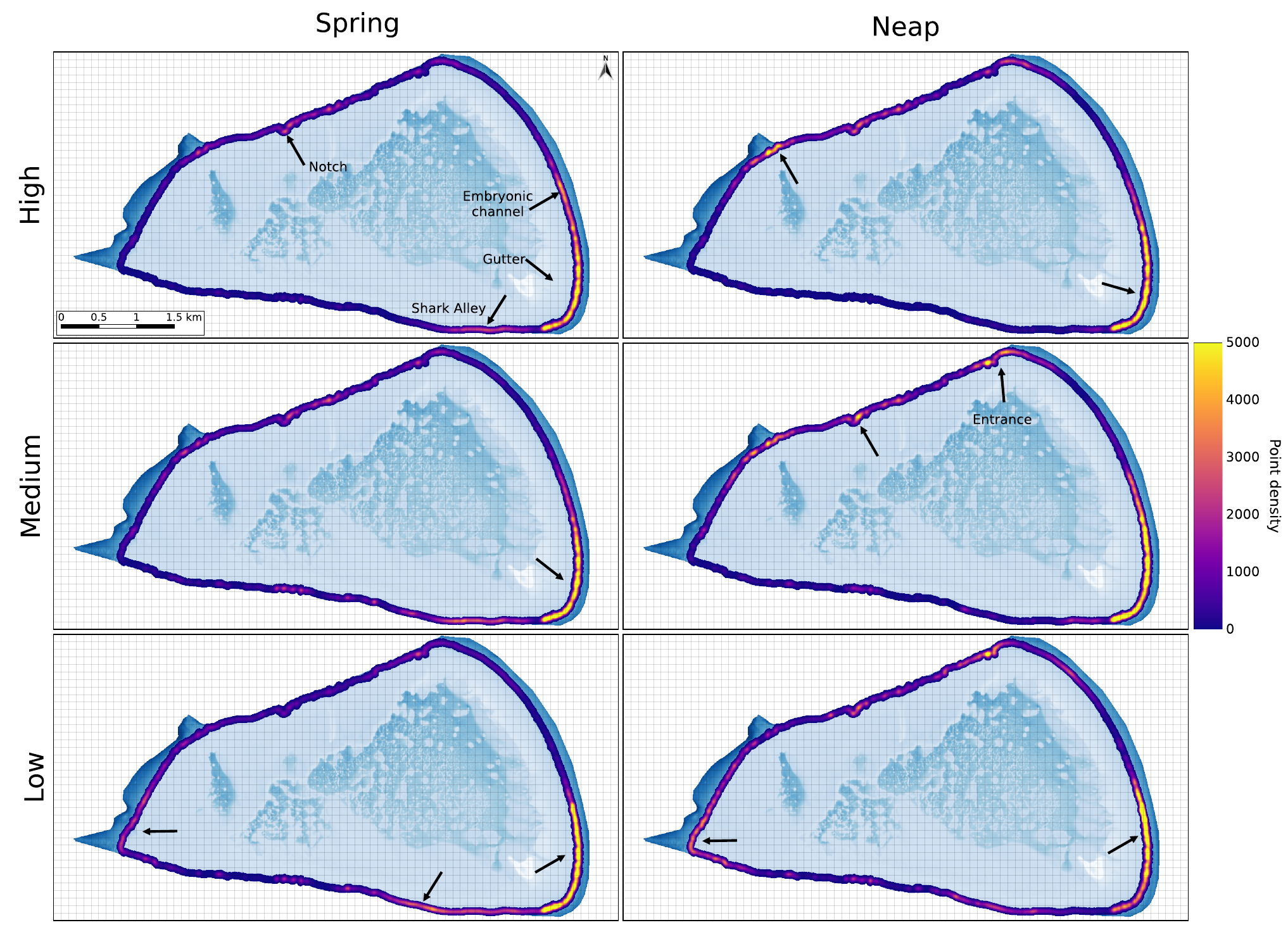}
\caption{Density maps showing the number of particles that cross out of the OTR area at each point on a 5 x 5 m grid. Areas of note are highlighted by arrows (see text). Top row shows the high resolution model output, centre show middle resolution and the bottom shows the low resolution model output. Left is the spring tidal period, right is the neap tidal period. Bathymetry is indicated by the blue colours using the LiDAR bathymetric map with darker colours indicating deeper water. Grid lines are at 100 m intervals for reference.}
\label{fig:outflow}
\end{figure}

\begin{figure}[ht]
\noindent\includegraphics[width=\textwidth]{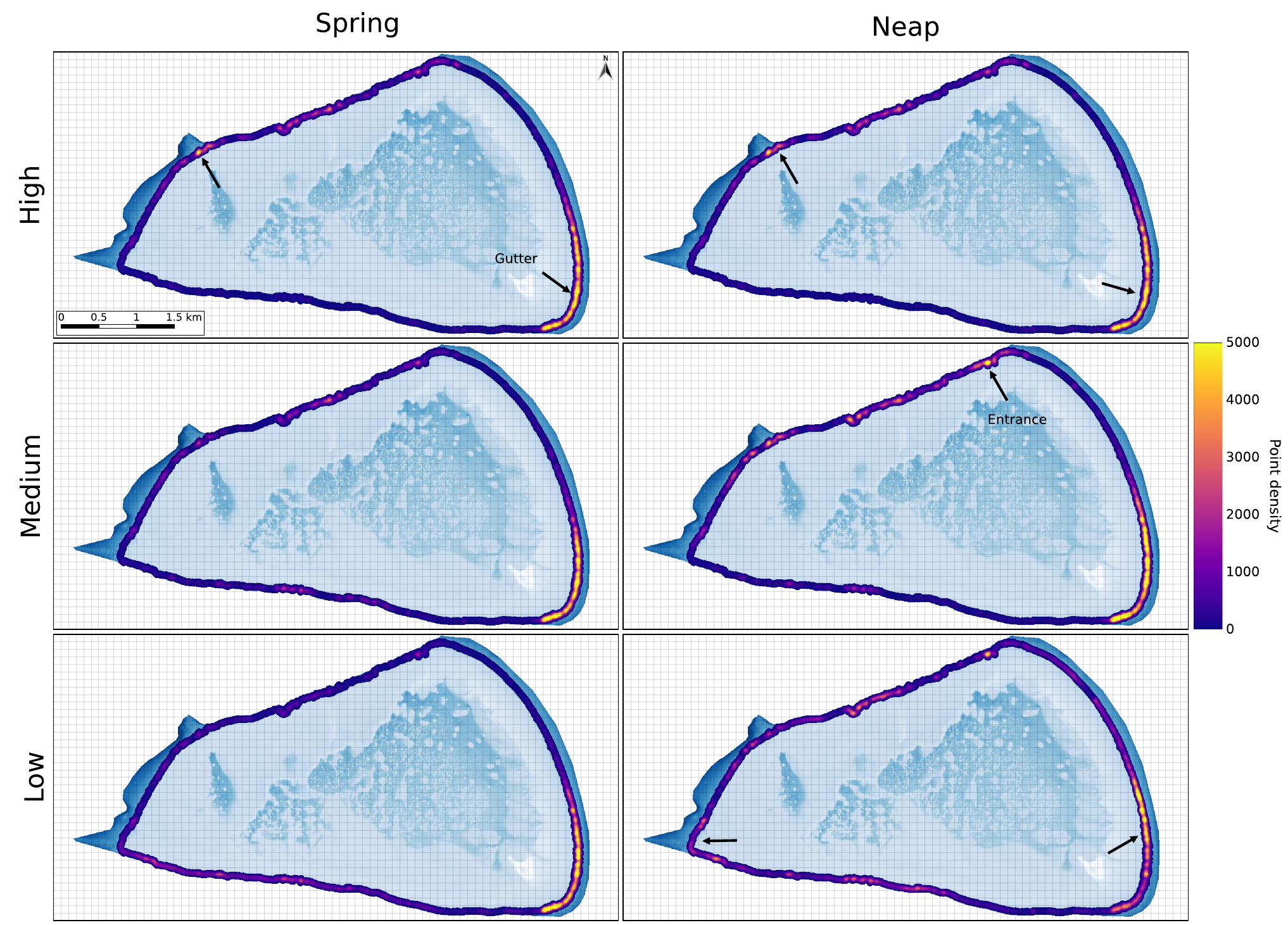}
\caption{Density maps showing the number of particles that cross into of the OTR area at each point on a 5 x 5 m grid. Areas of note are highlighted by arrows (see text). See Fig \ref{fig:outflow} for further details.}
\label{fig:inflow}
\end{figure}

The paths taken by particles show substantial differences in exit locations depending on resolution. Generally, the LR simulations show a diffuse patterns of exit locations around most of the reef rim, with a high concentration on the south-east corner (near `Gutter'). All models show few exit pathways across the north-east corner. As resolution increases, the exist locations become better defined. The HR simulation shows four distinct exit areas: the northern `Notch', the eastern edge of the reef (`embryonic channel' and `Gutter') and a channel on the southern margin (`Shark Alley') during spring conditions. In neap conditions, however, the exit locations shift slightly, with a new location to the west of `Notch' on the northern margin (Fig. \ref{fig:outflow}) The eastern edge of the reef remains a high density area, but `Shark Alley' is no longer a clear exit point. During spring tides, the MR model shows `Shark Alley' on the southern edge as a high density area, as well `Notch' and `Gutter'. However, `embryonic channel' is not as clear as the HR model. During neap tides there is a clear shift to the northern edge of the reef as exit points. A new high density region, `Entrance' appears (similar in location to \citet{Wilson1985-xm} point B) and `Notch' becomes higher density, as does a point to the west, similar to the HR model. However, like the HR model, the south-eastern corner remains the main exit location. Finally, the LR shows much less focused exit areas, but are somewhat similar to the high resolution model at spring tide, with the addition of a westerly exit point and lower density on the northern edge. At neap tide times, the exit points are similar to the medium resolution model, with the addition of the same high density area on the far western tip of the island as seen in the spring tide. Similar patterns are seen on entry points around the reef rim. However, notable exceptions include the lack of higher density in the area around `Shark Alley' and `embryonic channel' as seen in the HR model during the spring tide (Fig. \ref{fig:inflow}). For all other locations, the density maps for entry and exit are similar.  

\begin{table}[ht]
\label{tab:4}
\caption{Results form the Functional Data Analysis ANOVA tests represented as p-values for each model pair combination and the two tests used to compare the 16 groups of flushing rate time series per model.}
\begin{tabular}{lllll}
\multicolumn{1}{l|}{}         & \multicolumn{1}{c|}{L2N}          & \multicolumn{3}{c}{TRP}                                                        \\
\multicolumn{1}{l|}{\textbf{Neap}}     & \multicolumn{1}{c|}{}       & \multicolumn{1}{c}{ANOVA} & \multicolumn{1}{c}{ATS} & \multicolumn{1}{c}{WTPS} \\
\multicolumn{1}{l|}{LR-HR} & \multicolumn{1}{l|}{$p<0.001$} & $p<0.001$         & $p<0.001$       & $p<0.001$        \\
\multicolumn{1}{l|}{MR-HR} & \multicolumn{1}{l|}{$p<0.001$} & $p<0.001$         & $p<0.001$       & $p<0.001$        \\
\multicolumn{1}{l|}{MR-LR} & \multicolumn{1}{l|}{$p<0.001$} & $p<0.001$         & $p<0.001$       & $p<0.001$        \\
\multicolumn{5}{l}{\textbf{Spring}}                                                                                                                              \\
\multicolumn{1}{l|}{LR-HR} & \multicolumn{1}{l|}{$p<0.001$} & $p<0.001$         & $p<0.001$       & $p<0.001$        \\
\multicolumn{1}{l|}{MR-HR} & \multicolumn{1}{l|}{$p=0.052$} & $p=0.024$         & $p=0.017$       & $p=0.027$        \\
\multicolumn{1}{l|}{MR-LR} & \multicolumn{1}{l|}{$p<0.001$} & $p<0.001$         & $p<0.001$       & $p<0.001$       
\end{tabular}
\end{table}

\section{Discussion}

The results show very clear change of tidal hydrodynamics depending on the resolution of the model. In this experiment we have used the same model, same parameters and same bathymetric data and altered \emph{only} the numerical mesh. Any changes observed are therefore due to the resolution. However, this does not necessarily mean the higher resolution is more accurate or correct in any sense of those words. Here, we have demonstrated that the higher resolution model does replicate known tidal phenomenon on OTR; namely the asymmetric filling and emptying of the lagoon as measured by \citet{Wilson1985-xm}. It is possible that the lower resolution models could be tuned to replicate this aspect more faithfully, but this was not attempted in this work.

The change in resolution causes clear, and well defined, changes in flow patterns; this is more evident where wetting and drying is a major factor. The low resolution simulations do not capture the shape of OTR's reef around the island. This then inhibits the ponding effect and hence the lack of simulation of this physical phenomenon. In turn, this means additional analyses that depend on the flow, here lagoon flushing, but also analyses like coral larval connectivity, would be potentially flawed. Surprisingly, the medium resolution ($\sim$50 m) appeared to visually match the flow patterns well, but was significantly different in the secondary analysis of lagoon flushing -- both in terms of the rates and locations of outflow and inflow. A number of the locations where there is high outflow from the lagoon, the `Shark Alley` and `embryonic channel' are not highlighted in the MR and LR models. The lagoon flushing simulated here is much more rapid than at Scott Reef (Western Australia) which was simulated by \citet{Green2018-ev} using a virtual tracer method within a 35 m resolution model using a 10 m resolution bathymetric data set. Both One Tree and Scott Reef have similar tidal ranges (3.5 to 4 m) and semi-diurnal tides, but \citet{Green2018-ev} included wind and wave currents, which were excluded here to focus on tidal dynamics only. Simulating Scott Reef at extreme resolution or One Tree with wind and wave currents would allow a more direct comparison. 

All of the resolutions chosen here would all be considered as `high' resolution in many modelling domains; \citet{Saint-Amand2023-dg} stated that to model the Great Barrier Reef with any clarity of the reef matrix and rapid changes in bathymetry, a resolution of 250 - 500 m is required. Here, the lowest resolution was 250 m, so around that recommended by \citet{Saint-Amand2023-dg}. However, this resolution does not fully capture the tidal dynamics in and around a coral atoll like OTR. It is clear that wetting and drying processes are crucial to the simulation of the atoll and these are largely lost in the low resolution model. One Tree Island is around 200 m across and 400 m in the longest dimension so is not fully captured in the LR model. Once the model has sufficient resolution, the flow regime is more fully captured around the island. However, despite the similarities of the flow between the medium and high resolution models, the differences that do exist then produce statistically significant differences in secondary analysis that depend on the flow. This then explains previous studies that have examined fish egg and larval dispersal on OTR. \citet{Burgess2007-go} simulated the hydrodynamics around OTR, using a 300 m model, and captured the two large-scale eddies that form in the lee of the island on both the ebb and flood tides (`phase eddies'). However, their distribution of fish capture shows large numbers on the southern edge of OTR on both ebb and flood tides (their figure 3) which matches the `rolling eddies' simulated in the medium and high resolution models presented here. Similarly, \citet{Booth2000-fd} show that `Shark Alley` is a location of high larval settlement. \citet{Booth2000-fd} hypothesis that this is due to the connection between the lagoon and ocean, but other sites (`Entrance' and `Gutter') have similar geomorphological features, but were not sites of high recruitment. Here, we show that lagoon flushing pathways are concentrated around `Shark Alley' during the spring tide and `Gutter' during neap tide (Fig. \ref{fig:outflow}) . The high resolution tidal model can therefore explain this and also explain why \citet{Burgess2007-go} could not see these phenomenon in their 300 m resolution model. More widely, using different hydrodynamic models has shown to produce difference in particle tracking outcomes \citet{Choukroun2025-df}. Adding wave processes to the model presented here may change some of these conclusions and will be investigated in future. The predominant wave and wind direction on OTR is from the south-east \citet{Perris2024-ot}, so the main entry points on the eastern edge of the reef will likely be the same, but there may be fewer exit points on the eastern edge. We also anticipate the entry and exit points to become more diffuse when wind and waves are added due to the temporal variability in wind direction. 

The advent of digital twinning requires a thorough understanding of the physical processes that are resolvable within any numerical framework developed. Depending on the aim of the digital twin platform, for example simulating the impacts of coastal engineering, impacts of storms over a wide area, or examining changes in sea-surface height over an ocean, will require a different requirement in terms of resolution. This has been well known for numerical models for a long time, but is worth repeating as we move to digital twin platforms. These issues may be of larger consequence, depending on the pipeline used to create the digital twin. For example, a pipeline like that described in \citet{Chattopadhyay2024-bm}, which is a neural network trained on satellite data cannot, by definition, make predictions at a finer scale than that of the training data nor should it predict anything beyond the limits seen in the training data. To put it another way, we should not trust a prediction that is at a higher resolution than that of the training data or predicts events beyond those seen in the training data. A complication occurs when the training data is a mix of resolutions -- what do we take as the effective resolution of such a model? Adding different numerical models at a range of resolutions into the training data may mitigate these concerns as they will then encompass the range of reasonable (assuming the models are well validated and verified) outcomes. In addition, any numerical model will add a physical/chemical/biological mathematical framework, but as shown here, resolution is an important consideration in that training.

\section{Conclusions}
Model resolution has long known to be a source of error in numerical models of ocean and coastal processes. The creation of digital twins on either data alone or a mix of numerical models and data creates new issues, but the issue of model resolution still remains, particularly with the advancement of extremely high resolution data. Here, we examine what physical processes are revealed as we move from good resolution (250 m) to extreme resolution (5 m) within an unstructured mesh tidal model in a complex, coral reef atoll environment. We show that only the very high resolution can correctly simulate the ponding effect observed at this location. The differences in flow are largely due  to correctly simulating the wetting and drying around the atoll rim, which in turn dictate the pathways of water exchange on and off the atoll. At ultra-high resolution fine scale eddies (`rolling eddies') form as part of the ebb and flood cycle on the southern edge of the atoll. These in turn give a coherence to water parcels which can explain observations of fish egg and larvae distributions from previous studies. As we move from numerical modelling to digital twins, resolution of both the training data and any training model must be a consideration on what processes can be seen in the final outputs. 

\section{Data and videos}

Model inputs and processing scripts are available on github: \url{https://github.com/EnvModellingGroup/oti_resolution}.

Videos of model outputs are available on Figshare, \href{https://doi.org/10.6084/m9.figshare.30110200}{DOI: 10.6084/m9.figshare.30110200}

\section{Acknowledgements}

This project was undertaken on the Viking Cluster, which is a high performance compute facility provided by the University of York. We are grateful for computational support from the University of York High Performance Computing service, Viking and the Research Computing team. This work was supported by the Natural Environment Research Council (NERC) and the Adapting to the Challenges of a Changing Environment (ACCE) Doctoral Training Partnership [grant number NE/S00713X/1].

\clearpage
\bibliographystyle{cas-model2-names}
\bibliography{refs}

\end{document}